\documentclass{emulateapj}
\pdfoutput=1
\usepackage{graphicx}
\usepackage{gensymb}
\usepackage{amsmath}
\usepackage{natbib}
\usepackage{amsbsy}
\usepackage{subfig}
\usepackage{multirow}
\usepackage{textcomp}
\usepackage{setspace}
\usepackage{pslatex}
\usepackage{microtype}
 
\shorttitle{Properties of MHD sausage waves}
\shortauthors{Freij et al.}

\begin{document}
	
	\title{On the properties of slow MHD sausage waves within small-scale photospheric magnetic structures}
	
	\author{N. Freij$^{1}$, I. Dorotovi\v{c}$^{2}$, R. J. Morton$^{3}$, M. S. Ruderman$^{1,4}$, V. Karlovsk\'{y}$^{5}$ and R. Erd\'{e}lyi$^{1,6}$}
		
	\affil{$^{1}$ Solar Physics \& Space Plasma Research Centre (SP$^{2}$RC), School of Mathematics and Statistics, University of Sheffield, Hicks Building, Hounsfield Road, Sheffield, S3 7RH, United Kingdom}
	\email{n.freij@sheffield.ac.uk}
	
	\affil{$^{2}$Slovak Central Observatory, P. O. Box 42, SK-94701 Hurbanovo, Slovak Republic}
	\email{ivan.dorotovic@suh.sk}
		
	\affil{$^{3}$ Mathematical Modelling Lab, Northumbria University, Pandon Building, Camden Street, Newcastle upon Tyne, NE1 8ST, United Kingdom}
	\email{richard.morton@northumbria.ac.uk}
	
	\affil{$^{4}$ Space Research Institute (IKI), Russian Academy of Sciences, Moscow, Russia.}
	\email{m.s.ruderman@sheffield.ac.uk}
		
	\affil{$^{5}$ 	Hlohovec Observatory and Planetarium, Sl\'{a}dkovi\v{c}ova 41,
		SK-92001 Hlohovec, Slovak Republic}
	\email{astrokar@hl.cora.sk}
	
	\affil{$^{6}$ Debrecen Heliophysical Observatory, Research Centre for Astronomy and Earth Sciences, Hungarian Academy of Sciences, Debrecen, P.O.Box 30, H-4010, Hungary}
	\email{robertus@sheffield.ac.uk}
		
\begin{abstract}
	The presence of magneto-acoustic waves in magnetic structures in the solar atmosphere is well-documented.
    Applying the technique of solar magneto-seismology (SMS) allows us to infer the background properties of these structures.
	Here, we aim to identify properties of the observed magneto-acoustic waves and study the background properties of magnetic structures within the lower solar atmosphere.
	
	Using the Dutch Open Telescope (DOT) and Rapid Oscillations in the Solar Atmosphere (ROSA) instruments, we captured two series of high-resolution intensity images with short cadence of two isolated magnetic pores. 
	Combining wavelet analysis and empirical mode decomposition (EMD), we determined characteristic periods within the cross-sectional (\textit{i.e.,} area) and intensity time series.
	Then, by applying the theory of linear magnetohydrodynamics (MHD), we identified the mode of these oscillations within the MHD framework. 
		
	Several oscillations have been detected within these two magnetic pores. 
	Their periods range from 3 to 20 minutes.
	Combining wavelet analysis and EMD enables us to confidently find the phase difference between the area and intensity oscillations.
	From these observed features, we concluded that the detected oscillations can be classified as slow sausage MHD waves.
	Further, we determined several key properties of these oscillations such as the radial velocity perturbation, magnetic field perturbation and vertical wavenumber using solar magneto-seismology.
	The estimated range of the related wavenumbers reveals that these oscillations are trapped within these magnetic structures.
	Our results suggest that the detected oscillations are standing harmonics, and this allows us to estimate the expansion factor of the waveguides by employing SMS.
	The calculated expansion factor ranges from 4-12.
\end{abstract}
	
	
\section{Introduction}
\label{Intro}
	
	Improvements in space- and ground-based solar observations have permitted the detection and analysis of small-scale waveguide structures in the Sun's lower atmosphere.
	One such structure is a magnetic pore: a magnetic concentration with a diameter that ranges from $0.5$ to $6$ Mm with magnetic fields of $1$ to $3$ kG that typically last for less than a day \citep{1970SoPh...13...85S}.
	Magnetic pores are highly dynamic objects due to e.g., constant buffeting from the surrounding granulation in the photosphere.
	A collection of flows and oscillations have been observed within and around magnetic pores \citep{1999SoPh..187..389B,2002A&A...383..275H,2002A&A...395..249R,doretalb,SAO,2014A&A...563A..12D,freij2014,jess2015multiwavelength,2015A&A...579A..73M}.
	The major apparent difference between a sunspot and a magnetic pore is the lack of a penumbra: a region of strong and often very inclined magnetic field that surrounds the umbra. 
	
	It is important to understand which magnetohydrodynamic (MHD) waves or oscillatory modes can be supported in magnetic flux tubes in the present context.
	The reason for this is two-fold: it clarifies the observational signatures of each mode; and, whether that mode will manifest given the conditions of the local plasma. 
	Furthermore, absorption of the global acoustic \textit{p}-mode, and flux tube expansion will induce a myriad of MHD waves.	
	\citet{roberts} investigated how the slow mode may be extracted elegantly from the governing MHD equations, considering the special case of a vertical uniform magnetic field in a vertically stratified medium.
	The approach may, in principle, be generalized with non-uniform magnetic fields \citep{luna-cardozo} and, by taking into account non-linearity, background flows and dissipative effects.
	However, as we will show below, a first useful insight still can be made within the framework of ideal linear MHD applied to a static background.
	
	It is very difficult to directly (or often even indirectly) measure the background physical parameters (plasma-$\beta$ or density, for example) of localised solar structures.
	For the magnetic field, the  most common method is to measure the Stokes profiles of element lines in the lower solar atmosphere and then perform Stokes inversion in order to determine the magnetic field vectors.
	More recently, the development of solar magneto-seismology (SMS) has allowed the estimation of the local plasma properties which are generally impossible to measure directly \citep{Andries2009,Ruderman2009}.
	While this technique has been used for many years in the solar corona, only recently has it been applied to the lower solar atmosphere.
	For example, \citet{fujimura} accomplished this by observing and identifying wave behaviour in lower solar structures and interpreting the observed waves as standing MHD waves.
	A recent review on lower solar atmospheric application of MHD waves is given by e.g. \citet{banerjee,jess2015multiwavelength} and partially by \citet{2013SSRv..175....1M} in the context of Alfv{\'e}n waves.
		
	Extensive numerical modelling of wave propagation in small-scale flux tubes has been undertaken by \citet{khomenko,hasan2008dynamics,fedun2,fedun1,2011ApJ...730L..24K,2011AnGeo..29..883S,2012ApJ...755...18V,wedemeyer2012magnetic,Mumford2015}.
	These models are of localised magnetic flux tubes and the effect of vertical, horizontal or torsional coherent (sub) photospheric drivers mimicking plasma motion at (beneath) the solar surface on these flux tubes.
	It was found that extensive mode conversion may take place within flux tubes as well as the generation of slow and fast MHD modes or the Alfv{\'e}n mode that depended on the exact driver used.   
	
	\citet{vogler} and \citet{cameron}, using the MURaM code, simulated larger scale magnetic structures, including pores, to build up a detailed picture of the physical parameters (density, pressure and temperature) as well as flows in and around these structures, which has good observational agreement.
	
	\cite{doretala} observed the evolution of a magnetic pore's area for 11 hours in the sunspot group NOAA 7519 \citep[see][]{sobotka,doretalb}.
	They reported that the periodicities of the detected perturbations were in the range of 12-97 minutes and were interpreted as slow magneto-acoustic-gravity sausage MHD waves.
	\citet{morton}, using the Rapid Oscillations in the Solar Atmosphere (ROSA) instrument installed on the Dunn Solar Telecope (DST), also detected sausage oscillations in a solar pore. 
	The lack of Doppler velocity data made it difficult to conclude whether the waves were propagating or standing.
	The oscillatory phenomena were identified using a relatively new technique (at least to the solar community), known as Empirical Mode Decomposition (EMD).
	The EMD process decomposes a time series into Intrinsic Mode Functions (IMFs) which contain the intrinsic periods of the time series.
	Each IMF contains a different time-scale that exists in the original time series \citep[see][]{terradas}.
	This technique was first proposed by \citet{huang} and offers certain benefits over more traditional methods of period analysis, such as wavelets or Fourier transforms. 
	
	\cite{2014A&A...563A..12D} observed several large magnetic structures and analysed the change in time of the cross-sectional area and total intensity of these structures.
	Phase relations between the cross-sectional area and total intensity have been investigated by \citet[e.g.,][]{michal2013,moreels2013phase}.
	The phase difference found observationally was 0$\degree$, i.e., in-phase, which matches the phase relation for slow MHD sausage waves. 
	Further, these magnetic structures were able to support several oscillations with periods not too dissimilar to standing mode harmonics in an ideal case.    
	\cite{2015ApJ...806..132G} observed a magnetic pore within Active Region NOAA 11683, using high-resolution scans of multiple heights of the solar atmosphere using ROSA and Interferometric Bidimensional Spectrometer (IBIS) on the DST.
	They showed that sausage modes were present in all the observed layers which were damped whilst they propagated into the higher levels of solar atmosphere.
	The estimated energy flux that suggests the could contribute to the heating of the chromosphere.
	
	Standing waves are expected to exist in the lower solar atmosphere bounded by the photosphere and transition region \citep{mein,leibacher}. Numerical models also predict this behaviour \citep{zhugzhda1,erdelyi,malins}.
	Standing waves have been potentially seen in the lower solar atmosphere; using the Hinode space-borne instrument suite, \citet{fujimura} observed pores and inter-granular magnetic structures, finding perturbations in the magnetic field, velocity and intensity.
	The phase difference between these quantities gave an unclear picture as to what form of standing waves these oscillations were.
	Standing slow MHD waves have been detected in coronal loops with SoHO and TRACE (for reviews see, e.g., \citealp{wang2011standing,2012RSPTA.370.3193D}) and transverse (kink) oscillations have been detected in coronal loops \citep[e.g][for a review see \citealp{Andries2009,Ruderman2009}]{1999ApJ520880A,taroyan,oshea,2008ApJ...687L..45V}.
	Harmonics of a standing wave have potentially been seen in flare loops using ULTRACAM \citep[e.g.,][]{mathioudakis}.
	\citet{fleck} also reported the observation of standing waves in the lower solar chromosphere, by measuring the  brightness and velocity oscillations in Ca II lines. 
		
	In this article, we exploit phase relations between the area and intensity of two magnetic pores, in order to identify the wave mode of the observed oscillations.
	This information combined with the methods of solar magneto-seismology, allows us to determine several key properties of these oscillations and  of the magnetic structures themselves.
	Section \ref{DnA} details the observational data, its reduction and the analysis method.
	Section \ref{Wave} discusses the theory of the applicable MHD wave identification as well as the solar magneto-seismology equations used to estimate properties of the observed oscillations.
	Section \ref{res} contains the results of the data analysis, while Sect. \ref{conc} summarises.  

\section{Data and Method of Analysis}
\label{DnA}

	\begin{figure*}
		\centering
		\includegraphics[scale=0.5]{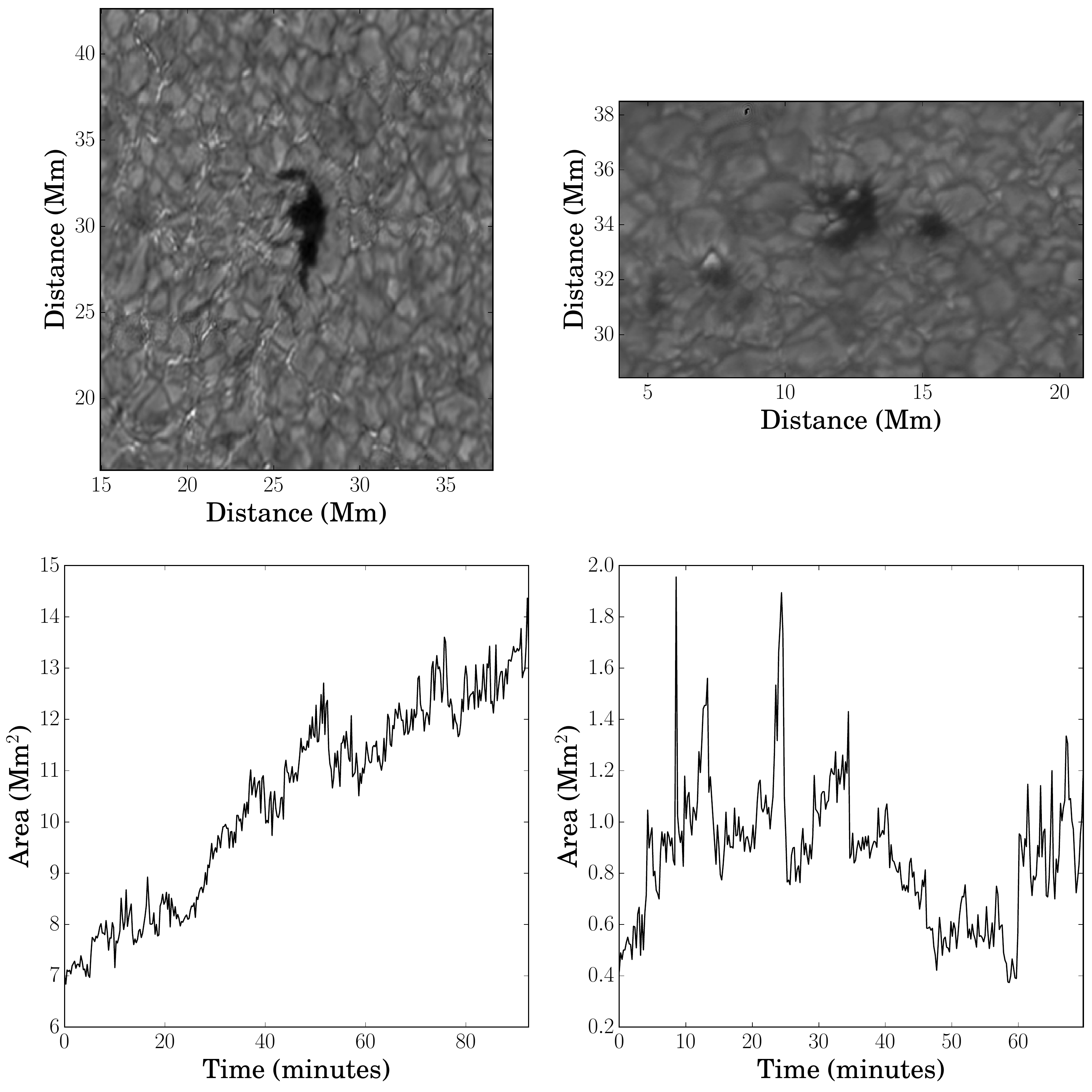}
		\caption{
				 The left column displays the magnetic pore observed by the DOT while the right column is the magnetic pore observed by the DST/ROSA.
				 The magnetic pores at the start of the observation sequence (\textit{Upper panels}).
				 The original (trended) cross-sectional time series for each pore throughout the observation sequence. (\textit{Lower panels}).
				 }
		\label{overview}
	\end{figure*}
	
	Two high-resolution datasets are investigated within this article.
	The first dataset was acquired using the Dutch Open Telescope (DOT) \citep{rutten}, located at La Palma in the Canary Islands. 
	The data were taken on $12$th August $2007$ with a G-band ($430.5$ nm) filter which samples the low photosphere and has a formation height of around $250$ km above the solar surface.
	The observation started at $08$:$12$ UTC and lasted for $92$ minutes with a cadence of $15$ seconds with a total field-of-view (FOV) of $60$ Mm by $40.75$ Mm.
	The DOT is able to achieve high spatial ($0.071''$ per pixel) resolution, due to the DOT reduction pipeline.
	It comes at a cost of temporal cadence which is decreased to $30$ seconds as data reduction uses speckle reconstruction \cite[]{1992A&A...261..321K}.
	Note that the DOT does not have an adaptive optics system.
	
	The second dataset was obtained on the $22$nd August $2008$ with the Rapid Oscillations in the Solar Atmosphere (ROSA) imaging system situated at the Dunn Solar Telescope (see \citealt{jess1} for details on experimental setup and data reduction techniques).
	Observation started at $15$:$24$ UTC, and data were taken using a $417$ nm bandpass filter with a width of $0.5$ nm.
	The $417$ nm spectral line corresponds to the blue continuum which samples the lower photosphere and the formation height of the filter wavelength corresponds to around $250$ km above the solar surface.
	It should be noted that this is an average formation height. 
	This is because the contributions to the line are from a wide range of heights and the lines also form at different heights depending on the plasma properties \citep{gband}.
		
	ROSA has the ability for high spatial ($0.069''$ per pixel) and temporal ($0.2$ s) resolutions.
	After processing through the ROSA pipeline the cadence was reduced to $12.8$ s to improve image quality via speckle reconstruction \citep{20764}.
	To ensure alignment between frames, the broadband time series was Fourier co-registered and de-stretched \citep{2007A&A...473..943J}.
	Count rates for intensity are normalised by the ROSA pipeline.

	The methodology of this analysis follows the one also applied by \citet{morton} and \cite{2014A&A...563A..12D}.
	The area of the pore is determined by summing the pixels that have intensity values less than $3\sigma$ of the median background intensity, which is a large quiet-Sun region. 
	This method contours the pore area well, but not perfectly, as the intensity between the pore and the background granulation is not a hard boundary.
	The top row of Figure \ref{overview} shows the magnetic pores at the start of the observation sequence, by DOT and ROSA, respectively. 
	Further, the output from the area analysis is shown in the bottom row for both magnetic pores.
	A strong linear trend can be observed for the DOT pore.
	The intensity time series was determined by total intensity of all the pixels within the pore.
	To search for periodic phenomena in the time series, two data analysis methods were used: wavelets and EMD.
	The wavelet analysis employs an algorithm that is a modified version of the tool developed by \citet{torrence}.
	The standard Morlet wavelet, which is a plane sine wave with the amplitude modulated by a Gaussian function, was chosen due to its high resolution in the frequency domain.
	The EMD code employed here is the one of \citet{terradas}.
	First, we de-trended each time series by linear regression followed by wavelet analysis to determine the periodicity of the oscillations as a function of time.
	Secondly, cross-wavelet is applied to calculate the phase difference between the area and intensity series as a function of time.
    Although it is possible to obtain a better visual picture of the phase relation between the two signals by using EMD, the results agreed with the cross-wavelet analysis when checked. 
		
\section{MHD wave theory}
\label{Wave}
	
	\subsection{The sausage mode}
	\label{Saus}
		
	We aim to identify MHD sausage modes and, as such, it is important to have a theoretical understanding of these modes.
    Assume, that a magnetic pore is modelled adequately by a cylindrical waveguide with a straight background magnetic field, i.e., $\textbf{B}_0=B_0\hat{\textbf{z}}$.
	We	note that, for reasons of clarity, in the following discussion the theory does not take into account gravitational effects on wave propagation.
	However, the influence of gravity may be important for wave propagation in magnetic pores, especially at the photospheric level where the predicted scale height is comparable to the wavelengths of observed oscillations.
    Therefore we should be cautious with the interpretations.
	The velocity perturbation is denoted as $\textbf{v}_1= (v_r,v_{\theta},v_z)$.
	From the theory of ideal linear MHD waves in cylindrical wave-guides, for the $m=0$ modes (here $m$ is the azimuthal wave number), i.e., for axi-symmetric perturbations, the equations determining $v_r$ and $v_z$ decouple from the governing equation for $v_{\theta}$.
	Hence, we will have magneto-acoustic modes described by $v_r$ and $v_z$ and the torsional Alfv\'en mode is described by $v_{\theta}$.
	We are interested in the slow magneto-acoustic mode in this paper, so we neglect the $v_{\theta}$ component.
	The	same applies to the component of the magnetic field in the $\theta$-direction. 
	The linear magneto-acoustic wave motion is then governed by the following ideal MHD equations,
	\begin{align}
		&&\rho_0 \frac{\partial v_r}{\partial t}=-\frac{\partial}{\partial r}
		\left(p_1+\frac{B_0b_z}{\mu_0}\right)+\frac{B_0}{\mu_0}\frac{\partial b_r}{\partial z},
		\label{eq:mom_r}\\
		&&\rho_0\frac{\partial v_z}{\partial t}=-\frac{\partial p_1}{\partial z},
		\label{eq:mom_z}\\
		&&\frac{\partial b_r}{\partial t}=B_0\frac{\partial v_r}{\partial z},
		\label{eq:mag_r}\\
		&&\frac{\partial b_z}{\partial t}=-B_0\frac{1}{r}\frac{\partial (rv_r)}{\partial r},
		\label{eq:mag_z}\\
		&&\frac{\partial p_1}{\partial t}=-\rho_0
		c_s^2\left(\frac{1}{r}\frac{\partial(rv_r)}{\partial r}+\frac{\partial v_z}{\partial z}\right),
		\label{eq:press1}\\
		&&\frac{\partial \rho_1}{\partial t}=-\rho_0\left(\frac{1}{r}\frac{\partial (rv_r)}{\partial r}+\frac{\partial v_z}{\partial z}\right).
		\label{eq:den}
	\end{align}
	Here, $p$ is the gas pressure, $\rho$ is the density and $\textbf{b} = (b_r,b_{\theta},b_z)$ is the perturbed magnetic field.
	We have assumed that the plasma motion is adiabatic.
	The subscripts $0$ and $1$ refer to unperturbed and perturbed states, respectively.
	
	Now, assume that the wave is harmonic and propagating and let $v_r=A(r)\cos(kz-\omega t)$.
	We then obtain the following equations for the perturbed variables,	
	\begin{align}
		&&\omega b_r=-B_0kv_r,\\
		&&\rho_0\left(\frac{v_A^2k^2}{\omega}-\omega\right)A(r)\sin(kz-\omega t)=\frac{\partial}{\partial r}\left(p_1+\frac{B_0b_z}{\mu_0}\right)
		&&\label{eq:n1}\\
		&&\rho_0\frac{\partial v_z}{\partial t}=-\frac{\partial p_1}{\partial z},
		\label{eq:n2}\\
		&& b_z=\frac{B_0}{\omega}\frac{1}{r}\frac{\partial (rA(r))}{\partial r}\sin(kz-\omega t),
		\label{eq:n3}\\
		&&\frac{\partial p_1}{\partial t}=c_s^2\frac{\partial\rho_1}{\partial t}=-\rho_0 c_s^2\left(\frac{1}{r}\frac{\partial(rv_r)}{\partial r}+\frac{\partial v_z}{\partial z}\right)
		\label{eq:n4}
	\end{align}
	Integrating Equation~(\ref{eq:n4}) with respect to $t$ and using (\ref{eq:n2}) gives
	\begin{align}
		&&p_1=c_s^2\rho_1=-\frac{\omega\rho_0
		c_s^2}{(c_s^2k^2-\omega^2)}\frac{1}{r}\frac{\partial
		(rA(r))}{\partial r}\sin(kz-\omega t).
		\label{eq:n5}
	\end{align}
	Comparing Equation (\ref{eq:n3}) to (\ref{eq:n5}) it can be noted that the magnetic field, $b_z$, and the pressure (density) are $180$ degrees out of phase.
	This depends on the sign of $c_s^2k^2-\omega^2$, which is assumed to be positive.
	Consideration of Equations~(\ref{eq:n1}), (\ref{eq:n3}) and (\ref{eq:n5}) leads to the conclusion that $v_r$ is $90\degree$ out of phase with $b_z$ and $-90\degree$ out of phase with $p_1$.
		
	The flux conservation equation for the perturbed variables gives the following relation,
	\begin{align}
		&&B_0S_1=-b_{1z}S_0,
		\label{eq:flux}
	\end{align}
	where $S$ refers to the cross-sectional area of the flux tube.
	We conclude that the perturbation of the area is out of phase with the perturbation of the  z-component of the magnetic field, hence, the area is in-phase with the fluctuations of the thermodynamic quantities.
	Perhaps more importantly, we re-write Equation~(\ref{eq:flux}) as 
	\begin{align}
		&&\frac{S_1}{S_0}=-\frac{b_{1z}}{B_0}.
		\label{eq:mag_area}
	\end{align}
	Hence, if we are able to measure oscillations of a pore's area, we can calculate the percentage change in the magnetic field due to these oscillations (assuming conservation of flux in the pore).
	This was previously suggested by \cite{2015ApJ...806..132G}.
	Exploiting this relation will allow a comparison to be made between the observed changes in pore area and the magnetic oscillations found from Stokes profiles (e.g. \citealp{Balthasar2000}).
	Further, as there are known difficulties with using the Stokes profiles, observing changes in pore area could provide a novel way of validating or refuting the observed magnetic oscillations derived from Stokes profiles.
	These simplified phase relations were confirmed in a more complicated case by e.g., \cite{michal2013} and \cite{moreels2013phase}, who also derived the phase relations for other linear MHD waves. 
	
	By measuring the change in pore area with time, we will also be able to estimate the amplitude of the radial velocity perturbation.
	The changes in area are related to changes in radius of the flux tube by
	\begin{align}
		&&\frac{S_1}{S_0}=\frac{2r_1}{r_0},
		\label{eq:area_rad}
	\end{align}
	where $r_0$ and $r_1$ are the unperturbed radius and perturbation of the radius, respectively, assuming the flux tube has a cylindrical geometry.
	Once a periodic change in radius is identified, the radial velocity of the perturbation can then be calculated using the following relation
	\begin{align}
		&&v_r=\frac{\partial r_1}{\partial t}=\frac{2\pi r_1}{P}.
		\label{eq:rad_vel}
	\end{align}
	
	Note, the term ``sausage mode'' was introduced for waves in magnetic tubes with a circular cross-section.
	The main property of these waves that distinguishes them from other wave modes is that they change the cross-sectional area.
	The cross sectional area of observed pores are typically non-circular.
	However, it seems to be reasonable to use the term sausage mode for any wave mode that changes the cross-sectional area.
	Several preceding papers have looked into non-circular, e.g., elliptic shapes, and found the effects to be marginal on the MHD waves within these tubes (see \citealt{2009A&A...494..295E} and \citealt{2011A&A...527A..53M}).

	\subsection{Period ratio of standing slow MHD wave}
	\label{stand}
	
	The period of a standing wave in a uniform and homogeneous flux tube is given by $P \approx 2L/nc_{ph}$, where $L$ is the tube length, $n$ is a integer determining the wave mode harmonic and $c_{ph}$ is the phase speed of the wave.
	This ratio is for ideal homogeneous tubes, however, this is not the case for the solar atmosphere from the photosphere to the transition region.
	\citet{luna-cardozo} modelled the effect of density stratification and expansion with height of the fluxtube on the ratio of the fundamental and first overtone periods for a vertical flux tube sandwiched between the photosphere and transition region.
	Their analysis studied the slow standing MHD sausage mode and assumed a thin flux tube with a small radial expansion with height. 
	They investigated two cases; case one is where the flux tube undergoes weak magnetic expansion with constant density, finding,  
	\begin{align}
		&&\frac{\omega_{2}}{\omega_{1}}= 2 - \frac{15}{2}\frac{\beta_{f}}{(6+5\beta_{f})\pi^{2}}(\Gamma-1),
		\label{mag_strat}
	\end{align}
	where $\omega_{i}$ is the period of specific harmonic or overtone (i.e., 1, 2), $\beta_{f}$ is the plasma-$\beta$ at the base of the flux tube and $\Gamma$ is the ratio of the radial size of the flux tube at the apex to the foot-point.
	Here, Equation (\ref{mag_strat}) is Equation (43) from \cite{luna-cardozo}.
	Case two is where the flux tube has density stratification but a constant vertical magnetic field, finding,
	\begin{align}
		&&\frac{\omega_{2}}{\omega_{1}}= \left[\frac{16\pi^{2} + \displaystyle\left(\ln\frac{1 - \sqrt{1 - \kappa_{1}}}{1 + \sqrt{1 - \kappa_{1}}}\right)^{2}}{4\pi^2 + \displaystyle\left(\ln\frac{1 - \sqrt{1 - \kappa_{1}}}{1 + \sqrt{1 - \kappa_{1}}}\right)^{2}} \right]^{1/2},
		\label{den_strat}
	\end{align} 
	where $\kappa_{1}$ is the square root of the ratio of the density at the top of the fluxtube to the density at the footpoint ($\kappa_{1} = (\rho_{apex}/\rho_{footpoint})^{0.5}$).
	Here, Equation (\ref{den_strat}) is Equation (40) from \cite{luna-cardozo}.
	Here, the upper end of the flux tube may well be the transition region while the footpoint is in the photosphere.
	It should be noted that the form of Equation (\ref{den_strat}) depends on the longitudinal density profile; here,a density profile where the tube speed increased linearly with height was used.
	This may or may not model a realistic pore and given the uncertainty of the equilibrium quantities this must be kept in mind in order to avoid over-interpretation.
	Equations(\ref{mag_strat}) and (\ref{den_strat}) modelling the frequency ratio of standing oscillations indicate that the ratio of the first harmonic to the fundamental will be less than two for fluxtube expansion while the density stratification could increase this value.
	Further, the thin fluxtube approximation is used to derive these equations.
	Obviously, in a real flux tube, both the density and magnetic stratification would be present at the same time and would alter the ratio.
	This is not accounted for at the moment.
	Further, Equations (\ref{mag_strat}) and (\ref{den_strat}) are independent of height which may limit the results as it has been suggested that the height to the transition region varies \citep{tian2009solar}.
	
\section{Results and Discussion}
\label{res}

	\begin{figure*}
		\centering
	    \subfloat{\includegraphics[width=0.45\textwidth]{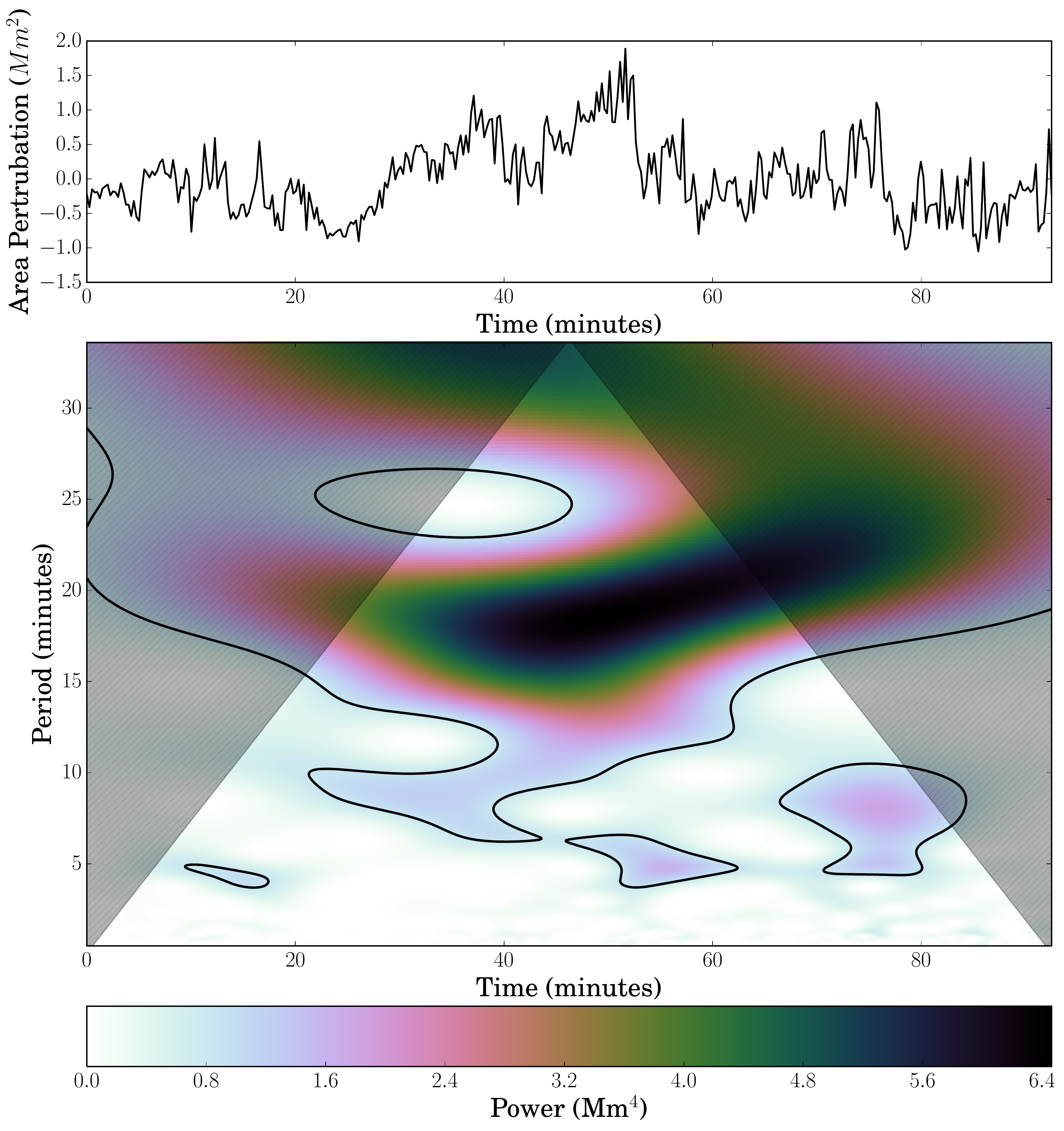}}
	    \subfloat{\includegraphics[width=0.45\textwidth]{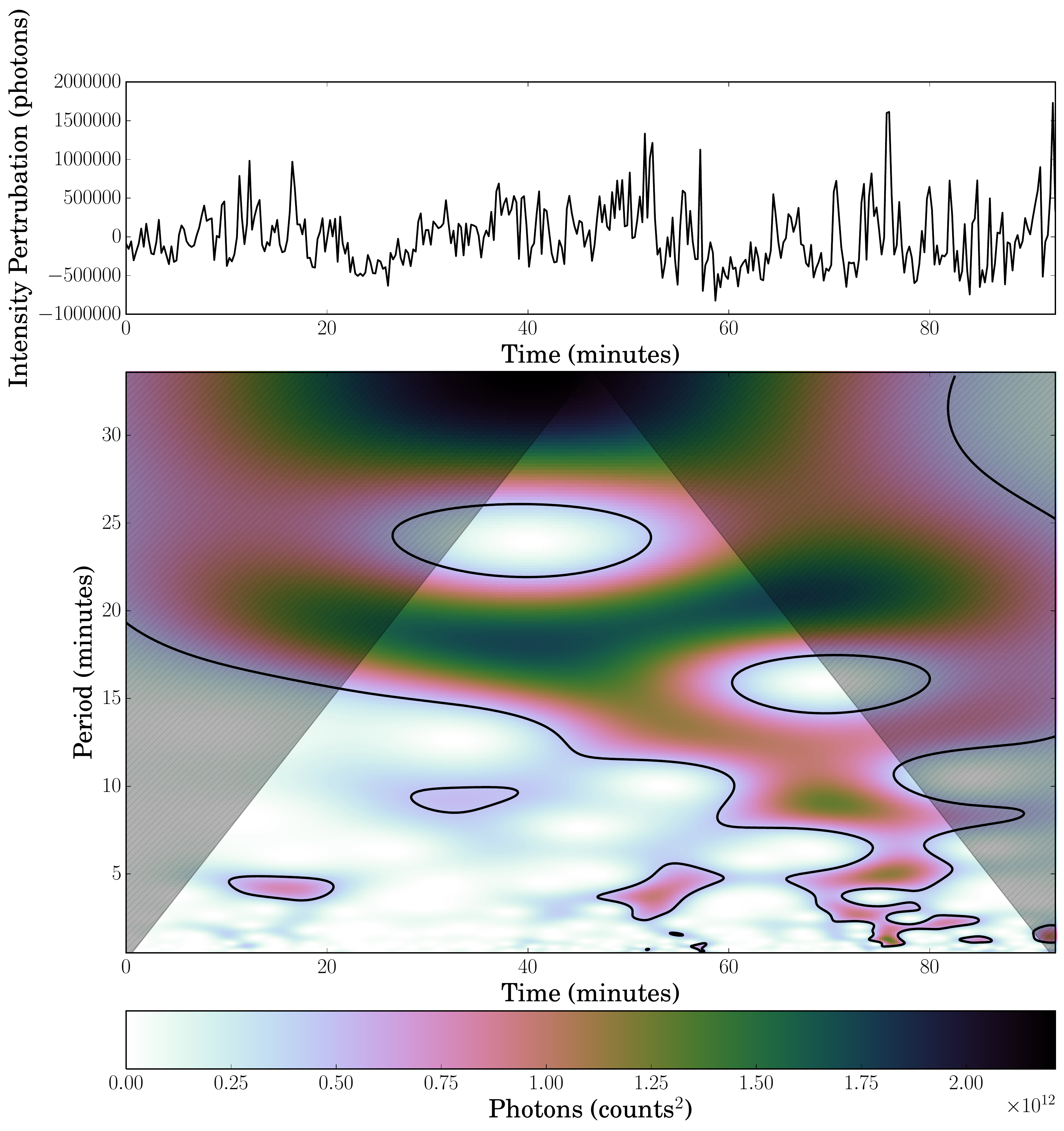}}
	    \caption{
		    	 Evolution of the area of the pore observed with DOT(\textit{Upper panels}).
	             The corresponding wavelet power spectrum for a white noise background.
	             The cone of influence is marked as the shaded region and the contour lines show the 95\% confidence level (\textit{Lower panels}).
	             }
	    \label{DOT_wls}
	\end{figure*}

	\begin{figure*}
		\centering
	    \subfloat{\includegraphics[width=0.45\textwidth]{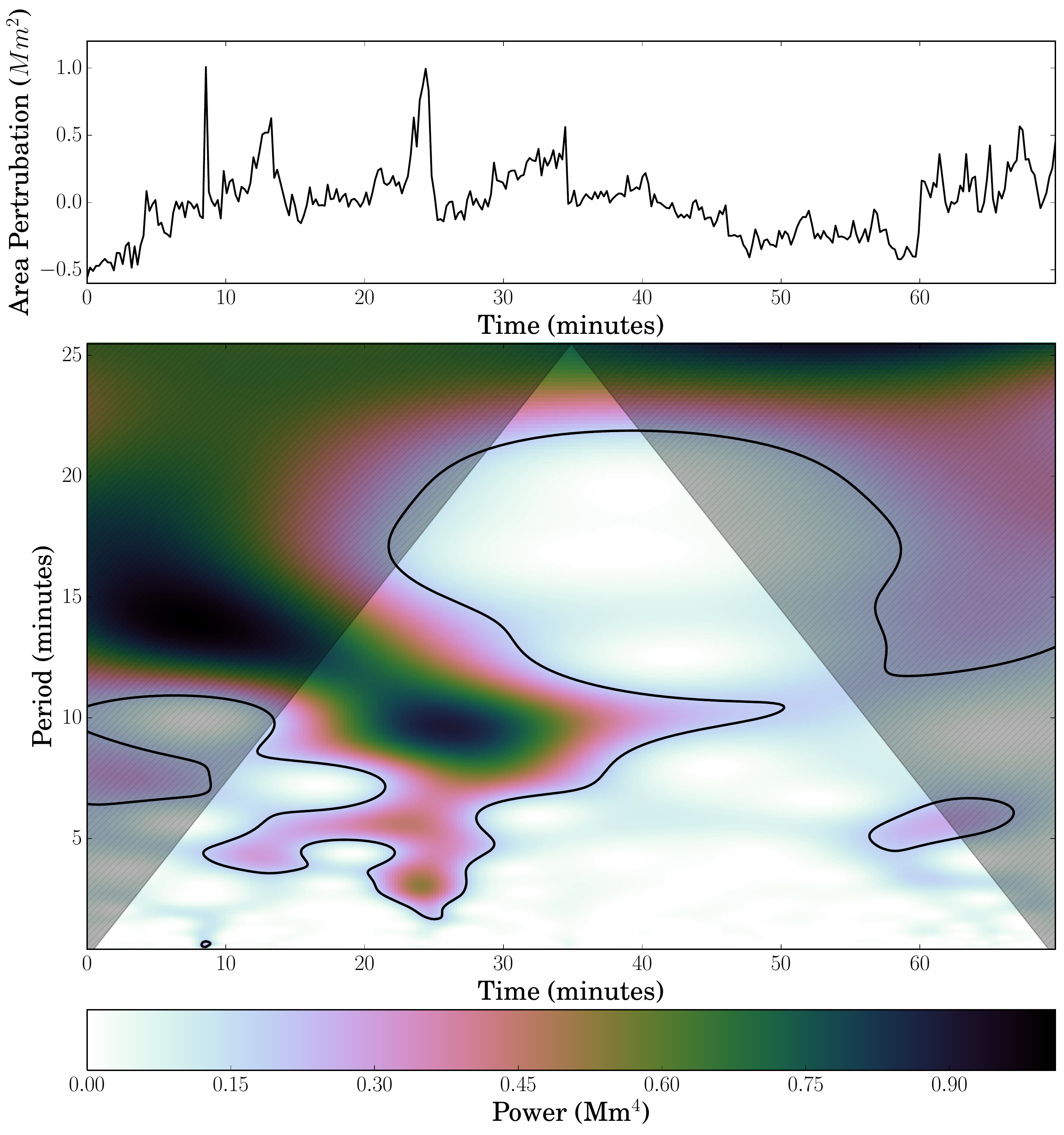}}
	    \subfloat{\includegraphics[width=0.45\textwidth]{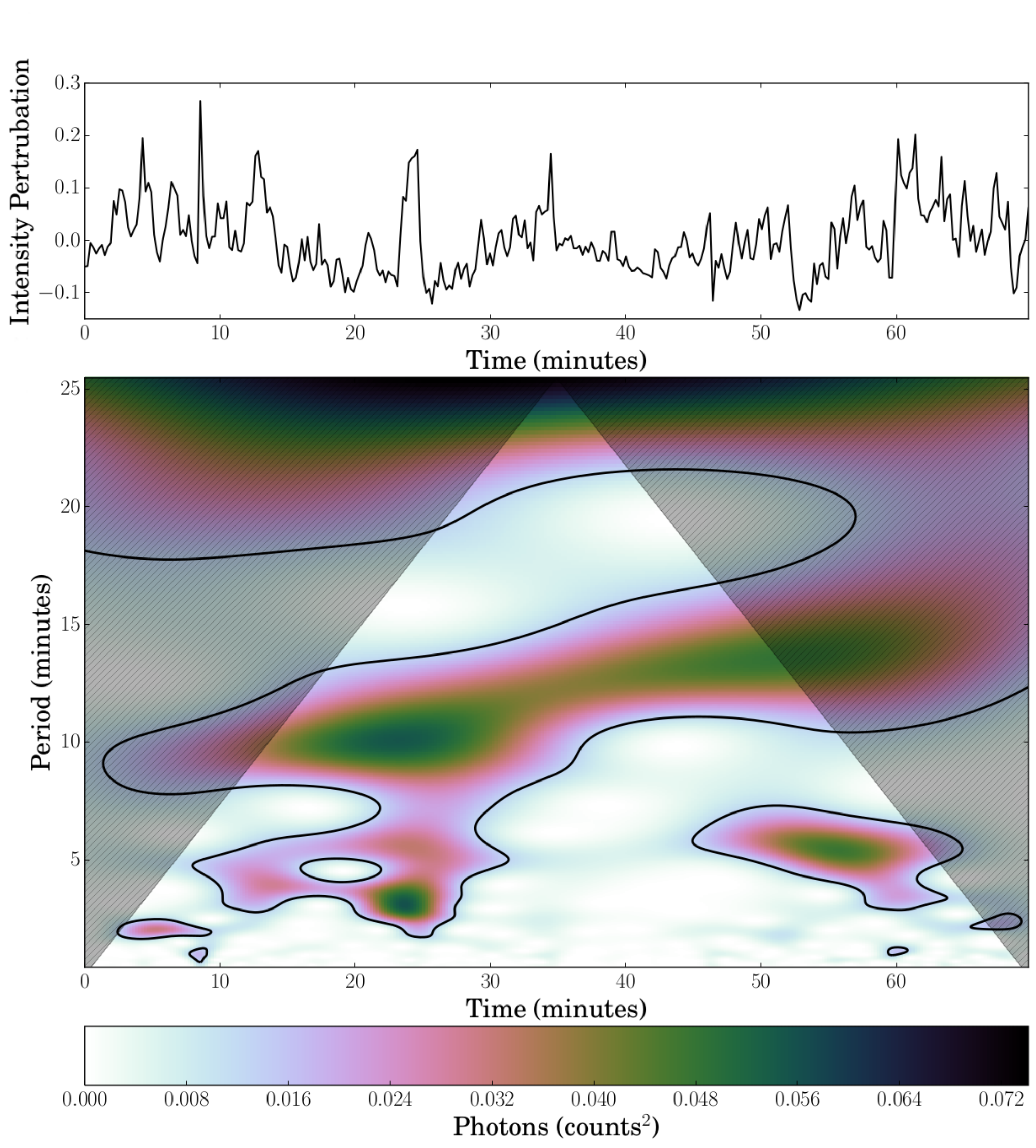}}
	    \caption{Same as Fig. \ref{DOT_wls} but for the pore observed with ROSA. Note that the intensity counts are normalised by the ROSA reduction pipeline \citep{jess1}.}
	    \label{ROSA_wls}
	\end{figure*}

	Figures \ref{DOT_wls} and \ref{ROSA_wls} show the results of wavelet analysis of the area and intensity time series for the DOT and DST telescopes, respectively.
	The original signal is displayed above the wavelet power spectrum, the shaded region marks the cone of influence (COI), where edge effects of the finite length of the data affect the wavelet transform results.
	The contours show the confidence level of 95\%. 
		
	\subsection{DOT Pore}
	
	There are four distinct periods found in the area time series of the pore; 4.7, 8.5, 20 and 32.6 minutes.
	The last period is outside the COI due to the duration of the time series, so it has been disregarded.
	It should be noted that periods of 8, 14 and 35 minutes have been observed in sunspots by \citet{kobanov}.
	This is important as pores and sunspots share a number of common features.
	The intensity wavelet shows 4 periods of oscillations; 4.7, 8.6, 19.7 and 35 minutes.
	These periods are similar, if not the same as the period of the area oscillations, which enables a direct comparison of the two quantities. 
	There is significant power that is co-temporal which can be observed in both the intensity and area wavelets.

	Using cross-wavelet in conjunction with the EMD allows the verification of the phase difference between the area and intensity signals for each period. 
	These methods show that the phase difference is very close to 0\degree, \textit{i.e.}, the oscillations are in-phase meaning that they are slow sausage MHD waves.	
	Further, the percentage change in intensity is also of the same order as reported in \citet{Balthasar2000} and \citet{fujimura}.
	This suggests, we are most likely observing the same oscillatory phenomena as these authors.

	We also have to be certain that any change in area we observe is due to the magneto-acoustic wave rather than a change in the optical depth of the plasma.
	\citet{fujimura} provide an insight into the expected differences between the phase of magnetic field and intensity oscillations due to waves or the opacity effect.
	They demonstrate that the magnetic field (pore area) should be in-phase (out-of-phase) with the intensity if the oscillations are due to changes in optical depth.
	We note that this is the same relationship expected for the fast magneto-acoustic sausage mode.
	Hence, the identification of the fast magneto-acoustic mode in pores may prove difficult with only limited datasets.

	The application of Equations (\ref{eq:area_rad}) and (\ref{eq:rad_vel}) require information about the amplitude of the area perturbation.
	This can be achieved using either an FFT power spectrum or the IMF's amplitude from the EMD analysis.
	Here, we use EMD for the amplitudes (which are time-average values) and they are $3.87\mathrm{x}10^5$ km$^2$, $3.61\mathrm{x}10^5$ km$^2$ and $5.90\mathrm{x}10^5$ km$^2$ for the oscillations with periods of 4.7, 8.5 and 20 minutes, respectively.
	It was not possible to find the amplitude of the largest period, as it did not appear in the EMD output.
	The values of area perturbation translate (using Equation \ref{eq:area_rad}) to 37, 34 and 56 km respectively for the amplitude of the radial perturbation.
	Note that the increase in radius is about $100$ km meaning the perturbation is only of the order of 1 pixel (at the DOT's resolution).

	Using the values above, allows us to calculate the radial velocity perturbation for each period (by means of Equation \ref{eq:rad_vel}).
	For the periods of 4.7, 8.5 and 20 minutes, we determine the radial velocity perturbation as 0.82, 0.42 and 0.29 km s$^{-1}$, respectively.
	The obtained radial speeds are very sub-sonic as the sound speed is $\approx$ 10 km s$^{-1}$ in the photosphere.
	They are, however, of the order of observed horizontal flows around pores.
	
	Further, it is also possible to estimate the percentage change in magnetic field expected from the identified linear slow MHD sausage modes.
	The percentage change in pore area, hence magnetic field, is found to be $$\frac{A_1}{A_0} = \frac{b_1}{B_0} \rightarrow 4-7\%.$$
	For another magnetic pore, the percentage change was found to be similar at 6\% \citep{2015ApJ...806..132G}.
	Let us now assume that the equilibrium magnetic field strength of the pore takes typical values of 1000-2000 G.
	Then, the amplitude of the magnetic field oscillations should be 40-140 G.
	The lower end of this estimated range of percentage change in magnetic field agrees well with the percentage changes in the magnetic field obtained using Stokes profiles by, for example, \citet{Balthasar2000} and \citet{fujimura}.
	However, the upper end of the range, i.e. $\sim 140$ G, appears twice as large as any of the previously reported periodic variations in magnetic field. 
	This apparent difference could be due to the spatial resolution of the magnetograms averaging out the magnetic field fluctuations.
	A summary of our findings can be found in Table \ref{tab:ampl}.
				
    \begin{table*}
        \centering
        \begin{tabular}{cccc|cccc}
            DOT & $r_{1}$ & $v_{r_1}$ & $\frac{b_{z1}}{B_{0}}$ & ROSA & $r_{1}$ & $v_{r_1}$ & $\frac{b_{z_1}}{B_{0}}$ \\ \hline \hline
            \begin{tabular}[c]{@{}l@{}}Period 1\\ 4.7 mins\end{tabular} & 37 km & 0.82 km s$^{-1}$ & 4.34\% & \begin{tabular}[c]{@{}l@{}}Period 1\\ 2-3 mins\end{tabular} & 69.1 km & 3.03 km s$^{-1}$ & 26.3\% \\
            \begin{tabular}[c]{@{}l@{}}Period 2\\ 8.4 mins\end{tabular} & 34 km & 0.42 km s$^{-1}$ & 4.04\% & \begin{tabular}[c]{@{}l@{}}Period 2\\ 5.5 mins\end{tabular} & 74.2 km & 1.41 km s$^{-1}$ & 28.2\% \\
            \begin{tabular}[c]{@{}l@{}}Period 3\\ 19.7 mins\end{tabular} & 56 km & 0.29 km s$^{-1}$ & 6.60\% & \begin{tabular}[c]{@{}l@{}}Period 3\\ 10 mins\end{tabular} & 117 km & 1.23 km s$^{-1}$ & 44.5\% \\
           \end{tabular}
           \caption{The properties of each observed period for the DOT and ROSA data respectively.
               $r_{1}$ is the radial perturbation, $v_{r1}$ is the velocity perturbation and $\frac{b_{z1}}{B_{0}}$ is the magnetic field perturbation.
               These quantities are determined by using Equations (\ref{eq:area_rad}) and (\ref{eq:rad_vel}).}
           \label{tab:ampl}
       \end{table*}
        
	Now, we estimate the wavelength (wavenumber) for each mode.
	An important fact needs to be remembered, \textit{i.e}, the velocity perturbation determined is radial, not vertical.
	Further, since the waveguide is strongly stratified, we define the wavelength as the distance between the first two nodes, which is the half wavelength of the wave.
	However, in this regime, the vertical phase speed of the slow sausage MHD wave is the tube speed, which is $c_T\approx4.5$ km s$^{-1}$ using typical values for the photospheric plasma \citep{1983SoPh...88..179E,evans}.
	For the periods of $4.7$, $8.5$ and $20$ minutes we obtain estimates of the wavelength (wavenumber) as $1269$ km ($4.95$ $\mathrm{x}10^{-6}$ m$^{-1}$), $2268$ km ($2.77$ $\mathrm{x}10^{-6}$ m$^{-1}$) and $5319$ km ($1.18$ $\mathrm{x}10^{-6}$ m$^{-1}$), respectively.
    Note that these wavelengths are larger than the scale height in the photosphere ($\approx 160$ km)  or the lower chromosphere.
	For the observed pore, it had an average radius, $a=1.5$ Mm, where $ka= 8, 5, 2$.
	See Table \ref{tab:wavelength} for a summary.
	
   \begin{table*}
       \centering
       \begin{tabular}{cccc|cccc}
           DOT & $\lambda_z$ & $k_z$ & $k_{z}a$ & ROSA & $\lambda_z$ & $k_z$ & $k_{z}a$ \\ \hline \hline
           \begin{tabular}[c]{@{}l@{}}Period 1\\ 4.7 mins\end{tabular} & 1269 km & 4.95 $\mathrm{x}10^{-6}$ m$^{-1}$ & 8 & \begin{tabular}[c]{@{}l@{}}Period 1\\ 2-3 mins\end{tabular} & 540-810 km & 7.76-12 $\mathrm{x}10^{-6}$ m$^{-1}$ & 4-6\\
           \begin{tabular}[c]{@{}l@{}}Period 2\\ 8.4 mins\end{tabular} & 2268 km & 2.77 $\mathrm{x}10^{-6}$ m$^{-1}$ & 5 & \begin{tabular}[c]{@{}l@{}}Period 2\\ 5.5 mins\end{tabular} & 1485 km & 4.2 $\mathrm{x}10^{-6}$ m$^{-1}$  & 2\\
           \begin{tabular}[c]{@{}l@{}}Period 3\\ 19.7 mins\end{tabular} & 5319 km & 1.18 $\mathrm{x}10^{-6}$ m$^{-1}$ & 2 & \begin{tabular}[c]{@{}l@{}}Period 3\\ 10 mins\end{tabular} & 2700 km & 2.33 $\mathrm{x}10^{-6}$ m$^{-1}$ & 1 \\
        \end{tabular}
        \caption{The wavelength (wavenumber) for each observed period for the DOT and ROSA data respectively.
            Here, $k=2\pi/\lambda$ and $\lambda=v/f$, where $k$ is the wavenumber, $\lambda$ is the wavelength, $v$ is the velocity and $f$ is the frequency.}
        \label{tab:wavelength}
    \end{table*}
    
	\subsection{ROSA Pore}
	
	There are four distinct periods found in the area time series of the pore observed by ROSA; 2-3, $5.5$, $10$ and $27$ minutes.
	All of these reported periods are at least at $95\%$ confidence level (or over).
	A few words about two of the periods have to be mentioned.
	Firstly, the power of the $2$-$3$ minute period is spread broadly and, as such, it is hard to differentiate the exact period.
	Secondly, the $10$-minute period slowly migrates to $13.5$ minutes as the time series comes to its end.
	The intensity wavelet shows four periods of oscillations; 2-3, 5.5, 10 and 27 minutes.
	For the pore observed by DOT, the oscillations found in the area and intensity data share similar periods.
	Also, there is another period that is below the $95\%$ confidence level for white noise at 1-2 minutes at the start of the time series.
	This is a similar behaviour as found for the DOT pore.

	We found that the phase difference between the area and intensity periods is 0\degree.
	This means, as before, that these oscillations are in-phase and are interpreted as signatures of slow sausage MHD waves.
	While we have chosen not to discuss the out-of-phase behaviour, there are small regions of 45$\degree$ phase difference that has been previously reported \citep{2014A&A...563A..12D}.
	This needs to be investigated in the future, as the authors are unaware of which MHD mode would cause this behaviour, however, it has been suggested that is due to noise within the dataset \citep{2015A&A...579A..73M}.
	As for the DOT pore, the same properties can be obtained for each period observed as within the ROSA pore and is summarized in Table \ref{tab:ampl} and \ref{tab:wavelength}.

	The amplitudes for the area oscillations are 2.29 $\mathrm{x}10^5$ km$^2$, 2.45 $\mathrm{x}10^5$ km$^2$ and 3.87 $\mathrm{x}10^5$ km$^2$ for periods of 2-3, 5.5 and 10 minutes, respectively.
	The $13.5$-minute period is found by the EMD process as well and has an amplitude which is the same as that of the $10$-minute period. 
	Again, it was not possible to find the amplitude of the largest period.
	These then, lead to the radial perturbation amplitude of 69.1, 74.2 and 117 km and the radial velocity perturbation as 3.03, 1.41 and 1.23 km s$^{-1}$, respectively.
	The increase in radius is around $100$ km meaning the perturbation is only of the order of 2 pixels (at ROSA's resolution).
	This means that for each part of the structure, its radius increases by 2 pixels.
	Once again, the radial velocity perturbations are found to be sub-sonic.
	
	The percentage change in the pore's area, and, thus the magnetic field is given by $$\frac{A_1}{A_0} = \frac{b_1}{B_0} \rightarrow 25 - 45\%.$$
	From the above relations we conclude that the size of the magnetic field oscillation is in the region of $200$-$400$ G.
	This is a substantial increase when compared to the measurements of the pore detected by DOT, as the amplitudes for these oscillations are of the same order but the cross-sectional area of the pore is an order of magnitude smaller.
	This suggests that the oscillation strength might be independent of the scale of the structure \citep{2014A&A...563A..12D}. 
	
	Once again, we determine the wavelength (wavenumber) for each period, using the tube speed as defined in the previous section.
	For the periods of $2-3$, $5.5$ and $10$ minutes we obtain estimates of the wavelength (wavenumber) as $540$-$810$ km ($7.76\mathrm{x}10^{-6}$ m$^{-1}$), 1485 km ($3.58\mathrm{x}10^{-6}$ m$^{-1}$) and 2.2 Mm ($2.85\mathrm{x}10^{-6}$ m$^{-1}$), respectively.
	For the observed pore radius, $a=0.5$ Mm, we obtain values of $ka = 2, 1.8, 1.5$ and $1.5$.
	
	\subsection{Standing Oscillations}

	\begin{table*}
	\centering
	\begin{tabular}{cc|cc}
		DOT Period (Mins) & Ratio ($P_{1}/P_{i}$) & Rosa Period (Mins) & Ratio ($P_{1}/P_{i}$)\\ \hline \hline
		\begin{tabular}[c]{@{}l@{}}\end{tabular} 8.5 mins & - & \begin{tabular}[c]{@{}l@{}}\end{tabular} 10 mins & - \\
		\begin{tabular}[c]{@{}l@{}}\end{tabular} 4.7 mins & 1.81 & \begin{tabular}[c]{@{}l@{}}\end{tabular} 5.5 mins & 1.81 \\
		\begin{tabular}[c]{@{}l@{}}\end{tabular}  & & \begin{tabular}[c]{@{}l@{}}\end{tabular} 2-3 mins & 3.3-5 \\
	\end{tabular}
		\caption{
                 The periods of oscillations as well as the harmonic ratios for the DOT and ROSA magnetic pore respectively.
				 The periods listed here exist at 95\% confidence level and are within the COI.
				 Periods greater than $10$ minutes have been neglected.}
		\label{harm_table}
	\end{table*}

	\begin{figure}
		\centering
		\includegraphics[width=0.45\textwidth]{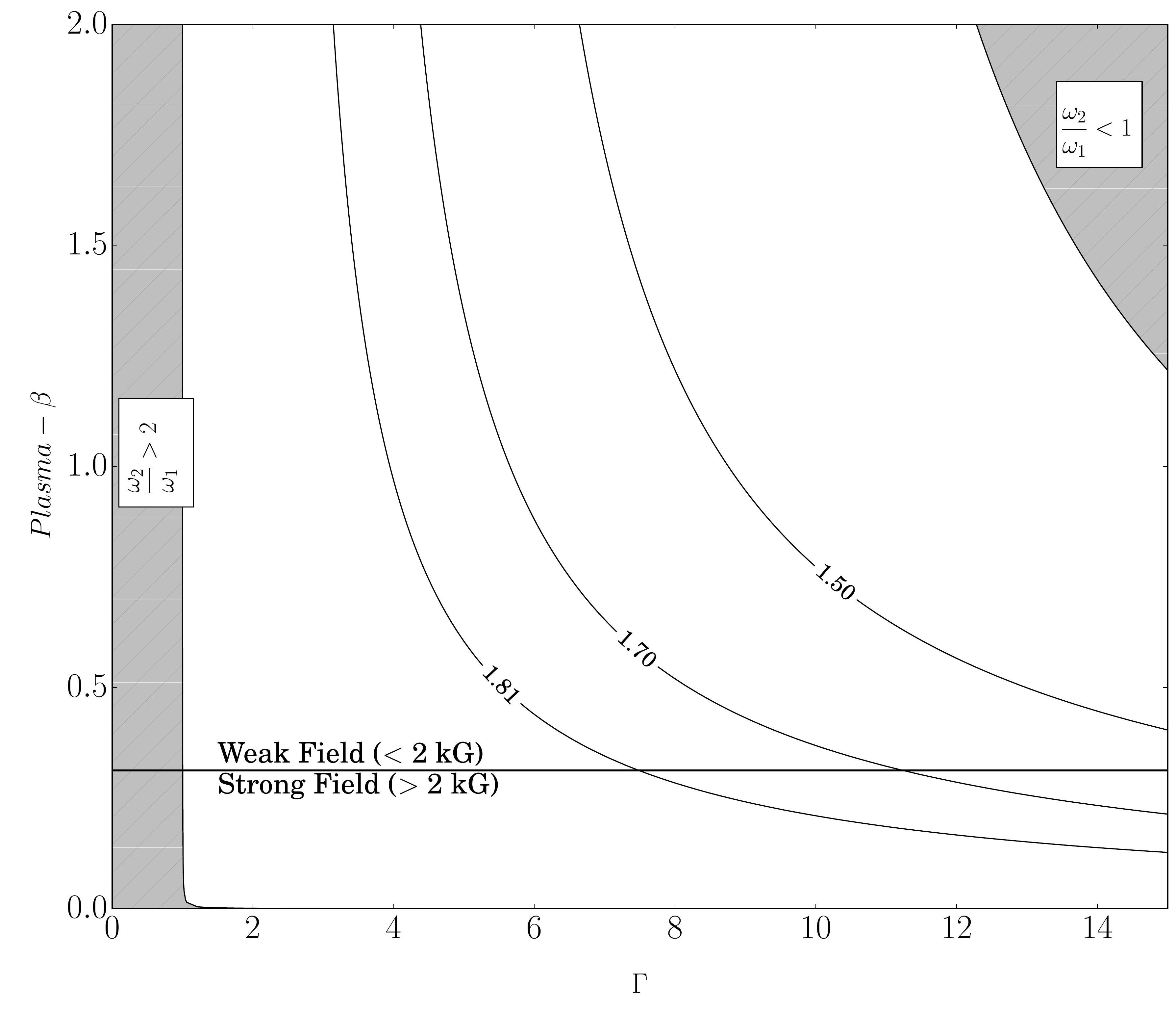}
	    \caption{The range of solutions for Equation (\ref{mag_strat}).
				 The threaded areas are where the period ratios are either less than one or greater than 2.
				 The horizontal line divides the image into a weak ($<2$ kG) and strong ($>2$ kG) field regions for the plasma-$\beta$.
				 The blue contour lines indicate observed period ratios for this paper and the values within \cite{2014A&A...563A..12D}.}
	    \label{fig:harm}
	\end{figure}		
		           
	With the important understanding that the observed waves are trapped, there is a possibility of them being standing ones.
	Assuming that the pore can be modelled as a straight homogeneous magnetic flux tube which does not expand with height, the sharp gradients (often modelled as discontinuities) of the temperature/density at the photosphere and at the transition region form a resonant cavity which can support standing waves \citep[see][]{fleck,malins}.  
		
	Calculating the harmonic periods ($P \approx 2L/nc_{ph}$, where $L$ is the distance between the boundaries ($2$ Mm), $n$ is the harmonic number), a fast MHD oscillation ($c_{p}\approx12$ kms$^{-1}$) would have a fundamental period $\sim$333 s, while the period of a slow MHD wave ($c_{p}\approx5.7$ kms$^{-1}$) would be $\sim$700 s.
	Other slow MHD sausage waves have been observed with phase speeds similar to this \citep{2015A&A...579A..73M}.
	The interpretation of the observed waves is that they are slow MHD sausage waves, which in the ideal homogeneous case is most similar to the observed results, however, it is still different by two minutes.
	Therefore, the basic assumption of an ideal homogeneous flux tube (constant $L$, constant $c_{ph}$ etc.) is inadequate to explain the results presented in this paper.
	There are several further considerations that need to be taken into account.
	From observations, many magnetic structures are not cylindrical or symmetrical and are often irregular in shape.
	Further to this, large-scale magnetic structures have been thought to be made up of either a tight collection of small-scale flux tubes or one large monolithic structure \citep[][and references within]{priest1984solar}.
	Also, these magnetic structures extend from the photosphere to the transition region which means that the plasma-$\beta$ will vary by an order of $2$ magnitude, which will change the dynamics of the MHD waves considerably.
	We have also ignored the effect of gravity (\textit{i.e.,} density stratification) \citep{2006A&A...458..975D,2011ApJ...743..164A}, as well as the equally important fact that flux tubes expand with height (\textit{i.e.,} magnetic stratification) which alters the ratio of the periods, i.e, $P_{1}/P_{2}\neq2$ \citep{luna-cardozo}. 
	All of these effects will further affect the wave dynamics inside flux tubes.
		
	Here, we will ignore periods greater than $10$ minutes as shown above, in the ideal homogeneous case, the largest period possible is $11.6$ minutes for MHD waves (with the above assumptions).
	Here, we will consider two effects: the effect of density stratification and magnetic expansion with height in the radial direction.
	For the first case; Equation (\ref{den_strat}) is calculated with typical density values from the VAL-III C model \citet{1981ApJS...45..635V} at the apex (transition region) and footpoint (photosphere) of the flux tube.
	The VAL-III C model is an estimation of a quiet-Sun region and the interior density ratio between photosphere and transition region of a flux tube need not necessarily differ greatly from that of the exterior atmosphere (see Figures 3 and 1 of \citealp{GFME13a} and \citealp{GFE14}, respectively). 
	The resulting value for the period ratio in this circumstance is 1.44 (density values are 2.727$\mathrm{x}10^{-7}$ and 2.122$\mathrm{x}10^{-13}$ g cm$^{-3}$ for the footpoint and apex respectively).
	Using the model given by \citet{Maltby1986}, which models a sunspot umbra, this period ratio is 1.38 (density values are 1.364$\mathrm{x}10^{-6}$ and 9.224$\mathrm{x}10^{-14}$ g cm$^{-3}$ for the footpoint and apex respectively).
	This does not correspond well to the results in this paper, but only for one previously reported result; a highly-dynamical non-radially uniform sunspot \citep{2014A&A...563A..12D}.
	The ratio is substantially smaller than what is detected here, which means the first harmonic should be at $\approx 5.9$ minutes.
	This model does not seem to be applicable to the observational results presented here. 
	The reason for this, the authors believe, is due to the effect of finite radius.
	The dispersion relation for slow MHD waves in a finite radial fluxtube, shows that the dispersion related to the finite tube radius increases the wave frequency.
	The shorter the wavelength, the stronger the dispersion effect is.
	Hence, the relative increase of the first overtone frequency due to the effect of finite radius is larger than that of the fundamental harmonic.
	This modifies the period of the first harmonic to be higher, which shifts the period ratio to be larger than values that are obtained theoretically in the thin tube approximation.
	
	Figure \ref{fig:harm} details the various solutions (\textit{i.e.,} period ratio) for Equation (\ref{mag_strat}) over a large range of plasma-$\beta$ and expansion ratio ($\Gamma$).
	It is difficult to estimate how much a flux tube expands with height, therefore, we explore the parameter space widely, taking $\Gamma$ of 0-15.
	The values for the plasma-$\beta$ is divided into strong ($\geq2$ kG) and weak ($\leq2$ kG) field regions, as the magnetic field of flux tubes hypothesised, will vary from $0.5$ kG to $4$ kG.
    The magnetic pores were observed before the launch of NASA's Solar Dynamics Observatory (SDO),
    so the best magnetic data comes from the Michelson Doppler Imager (MDI) instrument on-board NASA's Solar and Heliospheric Observatory (SOHO).
    As such, the magnetic field of these pores is hard to know precisely due to their small scale and MDI's large pixel size.
    However, ground-based observations of similar sized pores reveal magnetic fields ranging from $1$ kG to $2.5$ kG.
	The blue contour lines show the parameter space that matches the period ratios reported in this article and the ones in \citet{2014A&A...563A..12D}.
	For example, if the plasma-$\beta$ is around 1, the expansion factor for the three period ratios reported here are around $4$, $6$ and $9$.
    If we have plasma-$\beta$ $\ll 1$, the expansion ratio starts to increase rapidly. 

	Once again, this effect can be dominant when the flux tube expands too much, however, it is unlikely that a flux tube would expand by such a large amount.
	\citet{1982GApFD21237B}, for example, suggests that when the internal gas pressure exceeds the external gas pressure, the flux tube becomes unstable and this occurs when the flux tube expands greatly with height.
	
	For the cases presented in this paper, the flux tube has to expand four to six times to have a period ratio that is observed.
	In a number of numerical simulations that model these types of flux tubes, the magnetic field expands approximately 4-10 times which happens to be not too dissimilar to our findings \citep[see also][]{khomenko,fedun2,fedun1}.
	It should be noted that these estimates for expansion are for flux tubes with magnetic fields that have a field strength less than $2$ kG. 
	
	Unfortunately, as of yet, little is known about the source of the oscillations analysed in this paper.
	One possible origin of MHD sausage waves is suggested by e.g. \citet{khomenko} and \citet{fedun2}, where magneto-acoustic wave propagation in small-scale flux tubes was modelled using non-linear MHD simulations.
	One of the results of their simulations is that five-minute vertical drivers can generate a mixture of slow and fast sausage modes in localised magnetic flux tubes that propagate upwards.
	Furthermore, \citet{fedun1} model the effect of photospheric vortex motion on a thin flux tube, finding that vertex motions can excite dominantly slow sausage modes.
	However, these simulations need to be developed further, before we may comfortably link them to our assertions.
	
	Another potential source is from mode conversion that will occur at the lower region of the photosphere within sunspots and magnetic pores.
	For example, \cite{0004-637X-746-1-68}, modelled a background sunspot-like atmosphere and solving the non-linear ideal MHD equations for this system, found that the fast MHD wave will turn into a slow MHD sausage wave at the Alfv\'en-acoustic equipartition level (which is where the sound speed is equal the Alfv\'en speed) and the reverse is also true.
	The fast MHD wave to Alfv\'en conversion occurs higher up where there is a steep Alfv\'en speed gradient, as the fast MHD wave will reflect from this boundary.
	Below this level, the MHD waves are fast and above this level, slow MHD waves can be supported.
	This level occurs at approximately 200 km in their model.
	The observations used within this paper are thought to form at a height around 250 km.
	Further, sunspot umbra's are depressed in height and it would likely be the same for magnetic pores.
	These facts can offer an insight into the formation height of G-band since we believe that we are observing a primary slow acoustic mode modified by the magnetic field i.e., the slow MHD sausage wave.

	A word of caution: without LOS Doppler data, it is difficult to know whether the oscillations reported are standing or propagating.
	The data available for magnetic pores does not cover higher levels of the solar atmosphere such as the chromosphere or the transition region.
	The data presented here only represents a slice of the flux tube near the photosphere.
	Future work is needed to acquire simultaneous observations of magnetic pores in several wavelengths in order to sample the solar atmosphere at different heights.
	With detailed spectral images would allow other LOS quantities such as Doppler velocity and magnetic field to be measured.
	This way, the oscillations could be determined confidently as standing or propagating due to their different phase relations. 
	
\section{Conclusions}
\label{conc}
	
	The use of high-resolution data with short cadence, coupled with two methods of data analysis (wavelets and EMD), has allowed the observation of small-scale wave phenomena in magnetic waveguides situated on the solar surface.
	By studying the area and intensity perturbations of magnetic pores, it enables the investigation of the phase relations between these two quantities with the use of wavelets and EMD.
	The in-phase ($0^\circ$ phase difference) behaviour reveals that the oscillations observed are indicative of slow sausage MHD waves. 
	Further, with the amplitude of oscillations measured, several properties could be estimated; such as the amplitude of the magnetic field perturbation and radial speed of the perturbation.
	The scale of the magnetic field perturbation that are caused by slow MHD waves are of the order $10$\% and have radial speeds that are sub-sonic when compared to the sound speed at the photosphere.  
	With the MHD mode of these waves identified, the obtained vertical wavelength indicates that the flux tubes would have a strong reflection at the transition region boundary.
    Further indicating a chromospheric resonator. 
	Finally, the investigation of the period ratio of the oscillations suggests that the fundamental and first harmonic has been observed within these flux tubes.
	The period ratio observed coupled with magneto-seismology enabled an expansion factor to be calculated that was in very good agreement to values found in numerical models used for MHD wave simulations.
	
\begin{acknowledgements}
	The authors want to thank the anonymous referee for their comments as it has improved the clarity of this paper.
	The authors thank J. Terradas for providing the EMD routines used in the data analysis.
	Wavelet power spectra were calculated using a modified computing algorithms of wavelet transform original of which was developed and provided by C. Torrence and G. Compo, and is available at URL: http://paos.colorado.edu/research/wavelets/.
	The DOT is operated by Utrecht University (The Netherlands) at Observatorio del Roque de los Muchachos of the Instituto de Astrofsica de Canarias (Spain)funded by the Netherlands Organisation for Scientific Research NWO, The Netherlands Graduate School for Astronomy NOVA, and SOZOU.
	The DOT efforts were part of the European Solar Magnetism Network.
	NF, MSR and RE are grateful to the STFC and The Royal Society for the support received. 
    RE is also grateful to NSF, Hungary (OTKA, Ref. No. K83133).
\end{acknowledgements}

\bibliographystyle{apj}
\bibliography{saus_seis}

\begin{thebibliography}{}
\expandafter\ifx\csname natexlab\endcsname\relax\def\natexlab#1{#1}\fi

\bibitem[{{Andries} \& {Cally}(2011)}]{2011ApJ...743..164A}
{Andries}, J., \& {Cally}, P.~S. 2011, \apj, 743, 164

\bibitem[{Andries {et~al.}(2009)Andries, Van~Doorsselaere, Roberts, Verth,
  Verwichte, \& Erdélyi}]{Andries2009}
Andries, J., Van~Doorsselaere, T., Roberts, B., {et~al.} 2009, Space Science
  Reviews, 149, 3

\bibitem[{{Aschwanden} {et~al.}(1999){Aschwanden}, {Fletcher}, {Schrijver}, \&
  {Alexander}}]{1999ApJ520880A}
{Aschwanden}, M.~J., {Fletcher}, L., {Schrijver}, C.~J., \& {Alexander}, D.
  1999, \apj, 520, 880

\bibitem[{{Balthasar}(1999)}]{1999SoPh..187..389B}
{Balthasar}, H. 1999, \solphys, 187, 389

\bibitem[{Balthasar {et~al.}(2000)Balthasar, Collados, \&
  Muglach}]{Balthasar2000}
Balthasar, H., Collados, M., \& Muglach, K. 2000, Astronomische Nachrichten,
  321, 121

\bibitem[{Banerjee {et~al.}(2007)Banerjee, Erd\'{e}lyi, Oliver, \&
  O’Shea}]{banerjee}
Banerjee, D., Erd\'{e}lyi, R., Oliver, R., \& O’Shea, E. 2007, \solphys, 246,
  3

\bibitem[{{Browning} \& {Priest}(1982)}]{1982GApFD21237B}
{Browning}, P.~K., \& {Priest}, E.~R. 1982, Geophys. Astro. Fluid., 21, 237

\bibitem[{Cameron {et~al.}(2007)Cameron, Sch{\"u}ssler, V{\"o}gler, \&
  Zakharov}]{cameron}
Cameron, R., Sch{\"u}ssler, M., V{\"o}gler, A., \& Zakharov, V. 2007, \aap,
  474, 261

\bibitem[{{De Moortel} \& {Nakariakov}(2012)}]{2012RSPTA.370.3193D}
{De Moortel}, I., \& {Nakariakov}, V.~M. 2012, Royal Society of London
  Philosophical Transactions Series A, 370, 3193

\bibitem[{{D{\'{\i}}az} \& {Roberts}(2006)}]{2006A&A...458..975D}
{D{\'{\i}}az}, A.~J., \& {Roberts}, B. 2006, \aap, 458, 975

\bibitem[{{Dorotovi{\v c}} {et~al.}(2014){Dorotovi{\v c}}, {Erd{\'e}lyi},
  {Freij}, {Karlovsk{\'y}}, \& {M{\'a}rquez}}]{2014A&A...563A..12D}
{Dorotovi{\v c}}, I., {Erd{\'e}lyi}, R., {Freij}, N., {Karlovsk{\'y}}, V., \&
  {M{\'a}rquez}, I. 2014, \aap, 563, A12

\bibitem[{Dorotovi\v{c} {et~al.}(2008)Dorotovi\v{c}, Erd\'{e}lyi, \&
  Karlovsk\'{y}}]{doretala}
Dorotovi\v{c}, I., Erd\'{e}lyi, R., \& Karlovsk\'{y}, V. 2008, in Proc. IAU
  Symposium No. 247, ed. R.~Erd\'{e}lyi \& C.~A. Mendoza-Brice$\tilde{n}$o,
  Vol. 247 (Cambridge University Press), 351

\bibitem[{Dorotovi\v{c} {et~al.}(2002)Dorotovi\v{c}, Sobotka, Brandt, \&
  Simon}]{doretalb}
Dorotovi\v{c}, I., Sobotka, M., Brandt, P.~N., \& Simon, G.~W. 2002, \aap, 387,
  665

\bibitem[{{Edwin} \& {Roberts}(1983)}]{1983SoPh...88..179E}
{Edwin}, P.~M., \& {Roberts}, B. 1983, \solphys, 88, 179

\bibitem[{Erd{\'e}lyi {et~al.}(2007)Erd{\'e}lyi, Malins, T{\'o}th, \&
  De~Pontieu}]{erdelyi}
Erd{\'e}lyi, R., Malins, C., T{\'o}th, G., \& De~Pontieu, B. 2007, \aap, 467,
  1299

\bibitem[{{Erd{\'e}lyi} \& {Morton}(2009)}]{2009A&A...494..295E}
{Erd{\'e}lyi}, R., \& {Morton}, R.~J. 2009, \aap, 494, 295

\bibitem[{Evans \& Roberts(1990)}]{evans}
Evans, D., \& Roberts, B. 1990, \apj, 348, 346

\bibitem[{Fedun {et~al.}(2011{\natexlab{a}})Fedun, Shelyag, \&
  Erd{\'e}lyi}]{fedun2}
Fedun, V., Shelyag, S., \& Erd{\'e}lyi, R. 2011{\natexlab{a}}, \apj, 727, 17

\bibitem[{Fedun {et~al.}(2011{\natexlab{b}})Fedun, Shelyag, Verth,
  Mathioudakis, \& Erd{\'e}lyi}]{fedun1}
Fedun, V., Shelyag, S., Verth, G., Mathioudakis, M., \& Erd{\'e}lyi, R.
  2011{\natexlab{b}}, Ann. Geophys, 29, 1029

\bibitem[{Fleck \& Deubner(1989)}]{fleck}
Fleck, B., \& Deubner, F. 1989, \aap, 224, 245

\bibitem[{Freij {et~al.}(2014)Freij, Scullion, Nelson, Mumford, Wedemeyer, \&
  Erdélyi}]{freij2014}
Freij, N., Scullion, E.~M., Nelson, C.~J., {et~al.} 2014, The Astrophysical
  Journal, 791, 61

\bibitem[{Fujimura \& Tsuneta(2009)}]{fujimura}
Fujimura, D., \& Tsuneta, S. 2009, \apj, 702, 1443

\bibitem[{{Gent} {et~al.}(2014){Gent}, {Fedun}, \& {Erd{\'e}lyi}}]{GFE14}
{Gent}, F.~A., {Fedun}, V., \& {Erd{\'e}lyi}, R. 2014, \apj, 789, 42

\bibitem[{{Gent} {et~al.}(2013){Gent}, {Fedun}, {Mumford}, \&
  {Erd{\'e}lyi}}]{GFME13a}
{Gent}, F.~A., {Fedun}, V., {Mumford}, S.~J., \& {Erd{\'e}lyi}, R. 2013,
  \mnras, 435, 689

\bibitem[{{Grant} {et~al.}(2015){Grant}, {Jess}, {Moreels}, {Morton},
  {Christian}, {Giagkiozis}, {Verth}, {Fedun}, {Keys}, {Van Doorsselaere}, \&
  {Erd{\'e}lyi}}]{2015ApJ...806..132G}
{Grant}, S.~D.~T., {Jess}, D.~B., {Moreels}, M.~G., {et~al.} 2015, \apj, 806,
  132

\bibitem[{Hasan \& Van~Ballegooijen(2008)}]{hasan2008dynamics}
Hasan, S., \& Van~Ballegooijen, A. 2008, \apj, 680, 1542

\bibitem[{{Hirzberger} {et~al.}(2002){Hirzberger}, {Bonet}, {Sobotka},
  {V{\'a}zquez}, \& {Hanslmeier}}]{2002A&A...383..275H}
{Hirzberger}, J., {Bonet}, J.~A., {Sobotka}, M., {V{\'a}zquez}, M., \&
  {Hanslmeier}, A. 2002, \aap, 383, 275

\bibitem[{Huang {et~al.}(1998)Huang, Shen, Long, Wu, Shih, Zheng, Yen, Tung, \&
  Liu}]{huang}
Huang, N., Shen, Z., Long, S., {et~al.} 1998, Proceedings of the Royal Society
  of London. Series A: Mathematical, Physical and Engineering Sciences, 454,
  903

\bibitem[{Jess {et~al.}(2010)Jess, Mathioudakis, Christian, Keenan, Ryans, \&
  Crockett}]{jess1}
Jess, D., Mathioudakis, M., Christian, D., {et~al.} 2010, \solphys, 261, 363

\bibitem[{Jess {et~al.}(2015)Jess, Morton, Verth, Fedun, Grant, \&
  Giagkiozis}]{jess2015multiwavelength}
Jess, D., Morton, R., Verth, G., {et~al.} 2015, Space Science Reviews, 1

\bibitem[{{Jess} {et~al.}(2007){Jess}, {Andi{\'c}}, {Mathioudakis},
  {Bloomfield}, \& {Keenan}}]{2007A&A...473..943J}
{Jess}, D.~B., {Andi{\'c}}, A., {Mathioudakis}, M., {Bloomfield}, D.~S., \&
  {Keenan}, F.~P. 2007, \aap, 473, 943

\bibitem[{{Kato} {et~al.}(2011){Kato}, {Steiner}, {Steffen}, \&
  {Suematsu}}]{2011ApJ...730L..24K}
{Kato}, Y., {Steiner}, O., {Steffen}, M., \& {Suematsu}, Y. 2011, \apjl, 730,
  L24

\bibitem[{{Keller} \& {von der Luehe}(1992)}]{1992A&A...261..321K}
{Keller}, C.~U., \& {von der Luehe}, O. 1992, \aap, 261, 321

\bibitem[{Khomenko \& Cally(2012)}]{0004-637X-746-1-68}
Khomenko, E., \& Cally, P.~S. 2012, \apj, 746, 68

\bibitem[{Khomenko {et~al.}(2008)Khomenko, Collados, \& Felipe}]{khomenko}
Khomenko, E., Collados, M., \& Felipe, T. 2008, \solphys, 251, 589

\bibitem[{Kobanov \& Makarchik(2004)}]{kobanov}
Kobanov, N., \& Makarchik, D. 2004, Astron. Rep., 48, 954

\bibitem[{Leibacher {et~al.}(1982)Leibacher, Gouttebroze, \& Stein}]{leibacher}
Leibacher, J., Gouttebroze, P., \& Stein, R. 1982, \apj, 258, 393

\bibitem[{Luna-Cardozo {et~al.}(2012)Luna-Cardozo, Verth, \&
  Erd\'{e}lyi}]{luna-cardozo}
Luna-Cardozo, C., Verth, G., \& Erd\'{e}lyi, R. 2012, \apj, 748, 110

\bibitem[{Malins \& Erd\'{e}lyi(2007)}]{malins}
Malins, C., \& Erd\'{e}lyi, R. 2007, \solphys, 246, 41

\bibitem[{{Maltby} {et~al.}(1986){Maltby}, {Avrett}, {Carlsson},
  {Kjeldseth-Moe}, {Kurucz}, \& {Loeser}}]{Maltby1986}
{Maltby}, P., {Avrett}, E.~H., {Carlsson}, M., {et~al.} 1986, \apj, 306, 284

\bibitem[{Mathioudakis {et~al.}(2006)Mathioudakis, Bloomfield, Jess, Dhillon,
  \& Marsh}]{mathioudakis}
Mathioudakis, M., Bloomfield, D., Jess, D., Dhillon, V., \& Marsh, T. 2006,
  \aap, 456, 323

\bibitem[{{Mathioudakis} {et~al.}(2013){Mathioudakis}, {Jess}, \&
  {Erd{\'e}lyi}}]{2013SSRv..175....1M}
{Mathioudakis}, M., {Jess}, D.~B., \& {Erd{\'e}lyi}, R. 2013, \ssr, 175, 1

\bibitem[{Mein \& Mein(1976)}]{mein}
Mein, N., \& Mein, P. 1976, \solphys, 49, 231

\bibitem[{Moreels \& Van~Doorsselaere(2013)}]{moreels2013phase}
Moreels, M., \& Van~Doorsselaere, T. 2013, \AA, 551

\bibitem[{{Moreels} {et~al.}(2015){Moreels}, {Freij}, {Erd{\'e}lyi}, {Van
  Doorsselaere}, \& {Verth}}]{2015A&A...579A..73M}
{Moreels}, M.~G., {Freij}, N., {Erd{\'e}lyi}, R., {Van Doorsselaere}, T., \&
  {Verth}, G. 2015, \aap, 579, A73

\bibitem[{{Moreels} {et~al.}(2013){Moreels}, {Goossens}, \& {Van
  Doorsselaere}}]{michal2013}
{Moreels}, M.~G., {Goossens}, M., \& {Van Doorsselaere}, T. 2013, \aap, 555,
  A75

\bibitem[{Morton {et~al.}(2011)Morton, Erd\'{e}lyi, Jess, \&
  Mathioudakis}]{morton}
Morton, R.~J., Erd\'{e}lyi, R., Jess, D.~B., \& Mathioudakis, M. 2011, \apj,
  729, L18

\bibitem[{{Morton} \& {Ruderman}(2011)}]{2011A&A...527A..53M}
{Morton}, R.~J., \& {Ruderman}, M.~S. 2011, \aap, 527, A53

\bibitem[{Mumford {et~al.}(2015)Mumford, Fedun, \& Erdélyi}]{Mumford2015}
Mumford, S.~J., Fedun, V., \& Erdélyi, R. 2015, The Astrophysical Journal,
  799, 6

\bibitem[{O'Shea {et~al.}(2007)O'Shea, Srivastava, Doyle, \& Banerjee}]{oshea}
O'Shea, E., Srivastava, A., Doyle, J., \& Banerjee, D. 2007, \aap, 473, 13

\bibitem[{Priest(1984)}]{priest1984solar}
Priest, E. 1984, Solar magneto-hydrodynamics, Vol.~21 (Springer)

\bibitem[{Roberts(2006)}]{roberts}
Roberts, B. 2006, Phil. Trans. R. Soc. London. Ser. A, 364, 447

\bibitem[{{Roudier} {et~al.}(2002){Roudier}, {Bonet}, \&
  {Sobotka}}]{2002A&A...395..249R}
{Roudier}, T., {Bonet}, J.~A., \& {Sobotka}, M. 2002, \aap, 395, 249

\bibitem[{Ruderman \& Erd{\'e}lyi(2009)}]{Ruderman2009}
Ruderman, M., \& Erd{\'e}lyi, R. 2009, Space Science Reviews, 149, 199

\bibitem[{Rutten {et~al.}(2004)Rutten, Hammerschlag, Bettonvil, S{\"u}tterlin,
  \& De~Wijn}]{rutten}
Rutten, R., Hammerschlag, R., Bettonvil, F., S{\"u}tterlin, P., \& De~Wijn, A.
  2004, \aap, 413, 1183

\bibitem[{{Shelyag} {et~al.}(2011){Shelyag}, {Fedun}, {Keenan}, {Erd{\'e}lyi},
  \& {Mathioudakis}}]{2011AnGeo..29..883S}
{Shelyag}, S., {Fedun}, V., {Keenan}, F.~P., {Erd{\'e}lyi}, R., \&
  {Mathioudakis}, M. 2011, Annales Geophysicae, 29, 883

\bibitem[{{Simon} \& {Weiss}(1970)}]{1970SoPh...13...85S}
{Simon}, G.~W., \& {Weiss}, N.~O. 1970, \solphys, 13, 85

\bibitem[{Sobotka {et~al.}(1997)Sobotka, Brandt, \& Simon}]{sobotka}
Sobotka, M., Brandt, P.~N., \& Simon, G.~W. 1997, \aap, 328, 682

\bibitem[{Solanki(2003)}]{SAO}
Solanki, S.~K. 2003, \aapr, 11, 153

\bibitem[{Taroyan {et~al.}(2005)Taroyan, Erd{\'e}lyi, Doyle, \&
  Bradshaw}]{taroyan}
Taroyan, Y., Erd{\'e}lyi, R., Doyle, J., \& Bradshaw, S. 2005, \aap, 438, 713

\bibitem[{Terradas {et~al.}(2004)Terradas, Oliver, \& Ballester}]{terradas}
Terradas, J., Oliver, R., \& Ballester, J. 2004, \apj, 614, 435

\bibitem[{Tian {et~al.}(2009)Tian, Curdt, Teriaca, Landi, \&
  Marsch}]{tian2009solar}
Tian, H., Curdt, W., Teriaca, L., Landi, E., \& Marsch, E. 2009, \aap, 505, 307

\bibitem[{Torrence \& Compo(1998)}]{torrence}
Torrence, C., \& Compo, G. 1998, Bulletin of the American Meteorological
  Society, 79, 61

\bibitem[{Uitenbroek \& Tritschler(2006)}]{gband}
Uitenbroek, H., \& Tritschler, A. 2006, The Astrophysical Journal, 639, 525

\bibitem[{{Vernazza} {et~al.}(1981){Vernazza}, {Avrett}, \&
  {Loeser}}]{1981ApJS...45..635V}
{Vernazza}, J.~E., {Avrett}, E.~H., \& {Loeser}, R. 1981, \apjs, 45, 635

\bibitem[{{Verth} {et~al.}(2008){Verth}, {Erd{\'e}lyi}, \&
  {Jess}}]{2008ApJ...687L..45V}
{Verth}, G., {Erd{\'e}lyi}, R., \& {Jess}, D.~B. 2008, \apjl, 687, L45

\bibitem[{{Vigeesh} {et~al.}(2012){Vigeesh}, {Fedun}, {Hasan}, \&
  {Erd{\'e}lyi}}]{2012ApJ...755...18V}
{Vigeesh}, G., {Fedun}, V., {Hasan}, S.~S., \& {Erd{\'e}lyi}, R. 2012, \apj,
  755, 18

\bibitem[{V{\"o}gler {et~al.}(2005)V{\"o}gler, Shelyag, Sch{\"u}ssler,
  Cattaneo, Emonet, \& Linde}]{vogler}
V{\"o}gler, A., Shelyag, S., Sch{\"u}ssler, M., {et~al.} 2005, \aap, 429, 335

\bibitem[{Wang(2011)}]{wang2011standing}
Wang, T. 2011, \ssr, 158, 397

\bibitem[{Wedemeyer-B{\"o}hm {et~al.}(2012)Wedemeyer-B{\"o}hm, Scullion,
  Steiner, van~der Voort, de~La~Cruz~Rodriguez, Fedun, \&
  Erd{\'e}lyi}]{wedemeyer2012magnetic}
Wedemeyer-B{\"o}hm, S., Scullion, E., Steiner, O., {et~al.} 2012, Nature, 486,
  505

\bibitem[{W{\"{o}}ger {et~al.}(2008)W{\"{o}}ger, von~der L{\"{u}}he, \&
  Reardon}]{20764}
W{\"{o}}ger, F., von~der L{\"{u}}he, O., \& Reardon, K. 2008, Astron.
  Astrophys., 488, 375

\bibitem[{Zhugzhda \& Dzhalilov(1982)}]{zhugzhda1}
Zhugzhda, I., \& Dzhalilov, N. 1982, \aap, 112, 16

\end{thebibliography}

\end{document}